\documentclass[12pt]{amsart}
\usepackage{amssymb,amsmath,verbatim,amsfonts,amscd,amsbsy}
\usepackage[dvips]{graphicx,color}
\usepackage{graphics} \usepackage{epsfig} \usepackage{float}
\usepackage[mathscr]{euscript}
\newtheorem{theorem}{Theorem} \newtheorem{lemma}{Lemma}

\let\E\mathscr 
 \let\Bbb\mathbb
 
 \def\sup{\mathrm{sup}}
\def\inf{\mathrm{inf}}

\begin{document}

\title[$P = \lambda \mu$ and Length-Biased Sampling ]
{On the incidence-prevalence relation and length-biased sampling
}
\addtocounter{footnote}{1}
\footnotetext{Supported in part by
FQRNT and NSERC of Canada }

\author{ Vittorio Addona, Masoud Asgharian and David B. Wolfson}
\address{department of Mathematics and Computer Science, Macalester College, 1600 Grand Ave.,
St. Paul, MN, 55105} \email{addona@macalester.edu}
\address{department of mathematics and statistics, McGill University,
Burnside Hall, 805 Sherbrooke Street West, Montreal, Quebec,
CANADA H3A 2K6 } \email{asgharian@math.mcgill.ca}
\email{david@math.mcgill.ca}

\maketitle
\begin{center}
{\it Macalester College \ and \ McGill University}
\end{center}

\begin{abstract}
For many diseases, logistic and other constraints often render large
incidence studies difficult, if not impossible, to carry out. This
becomes a drawback, particularly when a new incidence study is
needed each time the disease incidence rate is investigated in a
different population. However, by carrying out a prevalent cohort
study with follow-up it is possible to estimate the incidence rate
if it is constant. In this paper we derive the maximum likelihood
estimator (MLE) of the overall incidence rate, $\lambda$, as well as
age-specific incidence rates, by exploiting the well known
epidemiologic relationship, prevalence = incidence $\times$ mean
duration ($P = \lambda \times \mu$). We establish the asymptotic
distributions of the MLEs, provide approximate confidence intervals
for the parameters, and point out that the MLE of $\lambda$ is
asymptotically most efficient. Moreover, the MLE of $\lambda$ is the
natural estimator obtained by substituting the marginal maximum
likelihood estimators for P and $\mu$, respectively, in the
expression $P = \lambda \times \mu$. Our work is related to that of
Keiding (1991, 2006), who, using a Markov process model, proposed
estimators for the incidence rate from a prevalent cohort study
\emph{without} follow-up, under three different scenarios. However,
each scenario requires assumptions that are both disease specific
and depend on the availability of epidemiologic data at the
population level. With follow-up, we are able to remove these
restrictions, and our results apply in a wide range of
circumstances. We apply our methods to data collected as part of the
Canadian Study of Health and Ageing to estimate the incidence rate
of dementia amongst elderly Canadians.
\end{abstract}


\section{Introduction}

In an incidence study, whose goal is to estimate a disease incidence
rate, a cohort of initially disease-free subjects is followed
forward in time. The subjects are monitored closely and for those
who develop the disease their approximate times of disease onset are
recorded. Often, as part of an incidence study, these diseased
subjects are followed until ``failure'' or censoring. The data
collected from such an incidence study may then be used to directly
estimate both the disease incidence rate and the survival function
for the time from onset to failure. The estimators of the incidence
rate and the survival function from such data are standard.

For many diseases, however, logistic and other constraints often render
large incidence studies difficult, if not impossible, to carry out.
This becomes a drawback, particularly when a new incidence study is needed
each time the disease incidence rate is investigated in a different population.
Nevertheless, by carrying out a prevalent cohort study with follow-up it is
possible to estimate the incidence rate if it is constant, thus avoiding the
problems associated with incidence studies. In this paper we derive the maximum
likelihood estimator (MLE) of the overall incidence rate, $\lambda$, as well
as age-specific incidence rates from data collected as part of a prevalent cohort
study with follow-up. We exploit the well known epidemiologic relationship,
prevalence = incidence $\times$ mean duration ($P = \lambda \times \mu$), to
suggest that the likelihood be derived as a function of the vector ($P$, $\mu$).
Once the MLE, $(\hat{P}, \hat{\mu})$ of $(P, \mu)$, is obtained, the MLE of
$\lambda = \frac{P}{\mu}$ follows by invariance. A similar approach may be used
to find the MLEs of age specific incidence rates. The asymptotic distributional
properties of the estimators may be obtained by modifying previous results for
the MLE of the survival function, based on survival data from a prevalent cohort
study with follow-up (see Section 4). It is comforting that the MLE  $\hat{\lambda}
= \frac{\hat{P}}{\hat{\mu}}$ is, therefore, also the natural ad hoc estimator of
$\lambda$.


In a medical setting, a prevalent cohort study with follow-up (Wang
1991) begins with the identification, from a sampled cohort, of
those with existing (prevalent) disease. The dates of onset for the
diseased are ascertained and the diseased subjects are followed
forward in time until failure or censoring. Other data collected
include the ages at the time of recruitment, the failure/censoring
times of the subjects who are followed, and covariates of interest
to the researchers. There are two main features of the data
collected from such studies. First, the dates of disease onset of
the prevalent cases do not include the dates of onset of those who
died prior to the start of the prevalent cohort study; we can only
speculate as to the existence of such subjects. Hence, direct use of
the observed dates of onset from a prevalence study, in contrast to
dates of onset from an incidence study, leads to underestimation of
the true incidence rate. Second, the observed failure/censoring
intervals are left-truncated and, if the underlying incidence
process is stationary, as is the assumption here, they are length
biased; those with longer survival intervals are more likely to be
observed (Wicksell 1925, Neyman 1955, Cox 1969,  Patil and Rao 1978,
and Vardi 1982, 1985). We address these difficulties in deriving the
MLE ($\hat{P}, \hat{\mu}$) and hence the MLE $\hat{\lambda}$, of the
overall incidence rate. We use a similar approach to the estimation
of age-specific incidence rates.

Keiding (1991, 2006) used a Markov process model to derive
carefully, the prevalence-incidence relationship, and proposed three
different scenarios which facilitate estimation of the (constant)
age-specific incidence rate when there are no follow-up data. In the
first scenario it is assumed that there is non-differential
mortality for the diseased and non-diseased. In Biering-Sorensen and
Hilden (1984) this assumption is likely to be tenable while in
Keiding et al. (1989), it is probably not, since the disease under
study is diabetes. In the second scenario, which Keiding invokes in
his 1989 paper, no assumption of non-differential mortality is made.
It is either assumed that the incidence rate is small and that the
difference between the intensities from the healthy and diseased
states to death is known or that the difference between these two
intensities is small and known. In the third scenario, it is assumed
that the joint relative intensity of the calendar time, age- and
duration-specific mortality is known. Under each scenario a
parametric assumption must be made, and in the last two scenarios
certain population parameters must be known. Therefore, these
estimators are strongly disease-specific and also dependent on the
availability of certain population level data. These assumptions are
needed to compensate for not having follow-up information. By
following-up the prevalent cases we are able to avoid these
assumptions. Our main assumption is that the underlying incidence
process is a stationary Poisson process, an assumption that Keiding
also makes. Stationarity of the incidence rate holds, roughly, for
many diseases: for example, amyotrophic lateral sclerosis (Sorenson
et al. 2002), certain types of cancers (Jemal et al. 2005), and
schizophrenia (Folnegovic and Folnegovic-Smalc 1992). It may not,
however, be tenable for an infectious disease.

Diamond and McDonald (1991) also considered incidence rate
estimation from prevalent-case data, with no follow-up, again
under different assumptions. Ogata et al. (2000) took an empirical
Bayes approach to the analysis of retrospective incidence data.
More recently Alioum et al. (2005) make HIV-AIDS-specific model
assumptions to estimate a general incidence rate. To our knowledge
there is no literature that provides a general framework for
maximum likelihood estimation of a constant underlying
incidence rate when one has access only to prevalent cohort
survival data with follow-up.

The rest of this paper is organized as follows:  In Section
\ref{notation} we provide a careful formulation of a prevalent
cohort study with follow-up, paying particular attention to survival
data. In Section \ref{charstat} we discuss the MLE for the
underlying incidence rate. In Section \ref{CIlambda}, we present the
asymptotic properties of the estimator, paving the way for
computation of an approximate confidence interval for the underlying
incidence rate. In Section \ref{age-specific} we extend our results
to include age-specific incidence rates. In Section
\ref{CSHAanalysis} we apply our methods to data collected as part of
the Canadian Study of Health and Aging (CSHA), in order to estimate
the underlying age-specific incidence rates of dementia amongst the
elderly in Canada.

\section{General setup and notation} \label{notation}

Let $X_1 , X_2 , ..., X_m$ be $m$ i.i.d. positive random variables
representing the survival times of individuals from onset of a
disease, say, to an end point of interest. Let the $X_i$'s have
survivor function $S(x)=P(X_i>x)$, cumulative distribution function
$F(x)$, and probability density function $f(x)$. Define $\mu$ to be
the mean survival time; that is, $\mu = \int_0^\infty S(x)dx$.
Suppose that $\tau_1, \tau_2, ..., \tau_m$ are the $m$ calendar
times of onset corresponding to $X_1, X_2, ..., X_m$ and let
$\tau^*$ be the calendar time of recruitment into a study.
Individual $i$ is observed in the study only if $X_i \geq \tau^* -
\tau_i$ and, therefore, for $i = 1, 2, ..., m$, $X_i$ is left
truncated with left truncation time $T_i = \tau^*-\tau_i$. Since the
onset times are random, the truncation times are random variables,
with distribution function denoted by $G$, and density $g$. Let $Y_1
, Y_2 , ..., Y_n$ be the \textit{observed} left truncated lifetimes,
with $n \leq m$. That is, $P(Y_i>x)=P(X_i>x|X_i>T_i)$. We borrow
terminology from renewal theory and write $Y_i
= Y_i^{bwd} + Y_i^{fwd}$, where $Y_i^{bwd}$ is the time from onset
to recruitment into the study or the ``backward recurrence time'',
and $Y_i^{fwd}$ is the time from recruitment to failure, or the
``forward recurrence time''. Also, let $F_{LB}$ represent the
distribution of the $Y_i$'s, where the subscript, $LB$, is used to
indicate that the $Y_i$'s are length-biased.

Suppose that individual $i$ has censoring time $C_i^* = Y_i^{bwd}
+ C_i$, where $C_i$, which we call the ``residual censoring
time'', is the time from recruitment until the individual is
censored. We assume that P($C_i^* > T_i$) = 1 (see Wang 1991) and
we thus observe only $\min(C_i^*,Y_i)$. Often, however, the
backward recurrence times are fully observed, and we assume this
to be the case here, so that the observed data are:
$(Y_i^{bwd}, Y_i^{obs}, \delta_i) \ i = 1,2, ..., n$
where $Y_i^{obs} = min(Y_i^{fwd}, C_i)$ and $\delta_i =
I[Y_i^{fwd} \leq C_i]$ indicates whether subject $i$ has been
followed until failure. Since $C_i^*$ and $Y_i$ share $Y_i^{bwd}$,
the full survival times, $Y_i$, are informatively censored (Vardi
1989). It is still reasonable in many cases, however, to assume
that $C_i$ is independent of both $Y_i^{fwd}$ and $Y_i^{bwd}$,
since independence between $C_i$ and $Y_i^{fwd}$ corresponds to
the usual random censoring assumption.

In summary, we differentiate between the {\it potential} failure times, $X_{i}$,
some of which will not be observed because of left truncation, and the observed
failure/censoring times ($Y_i^{bwd}$ + $Y_{i}^{obs}$); these are event times of
the ``long survivors''.

\section{Point estimation of the incidence rate} \label{charstat}

Under stationarity we assume the underlying incidence process is a
Poisson process with constant intensity $\lambda(t) \equiv \lambda$.
Hence, the truncation time distribution, $G$, is uniform,
conditional on the number of incident times in $(0,\tau^*)$
(Asgharian, Wolfson, and Zhang 2006).

We begin by deriving the MLE of $\lambda$. We then derive the MLE of
the age-specific incidence rates. The approach depends on the
well-known relationship, $P = \lambda \times \mu$ where $P$ is the
time-independent point prevalence, $\lambda$ is the time-independent
underlying incidence rate, and $\mu$ is the mean duration of the
disease (see Keiding 1991).

Let a random number of $N$ prevalent cases be observed from a large
group of $s$ individuals selected from screening. Fixing $N$ at $n$,
the realized number of prevalent cases, Asgharian, M'Lan, and
Wolfson (2002) and Asgharian and Wolfson (2005) derived the
unconditional NPMLE, $\hat{S}$, of $S$ under the assumption of
stationarity. They established its asymptotic properties and those
of the NPMLE, $\hat{\mu} = \int_0^\infty \hat{S}(x)dx$. Conditioning
on $N=n$, the likelihood of the data is
\begin{equation} \label{condLikeN}
L = L(S,P) = f(\text{data} , n ; S , P).
\end{equation}
Write $L(S,P)$ as $L(S | n)L(P) = f(\text{data} ; S | n) f(n ; P)$,
by sufficiency of $N$ for $P$. Now, in practice, the data of a
prevalent cohort study with follow-up are collected in two stages.
In stage 1, a binary, 0-1, random variable, say $\xi$, is measured
on each randomly selected subject to ascertain if the subject has
experienced initiation of the disease. In stage 2, we observe the
triple $(Y_i^{bwd}, Y_i^{obs}, \delta_i) \ i = 1,2, ..., n$ 
on diseased subjects, indicated by $\xi=1$.
The following tree diagram depicts our sampling scheme:
\begin{figure}[!h]
\centering
\includegraphics[scale=0.70]{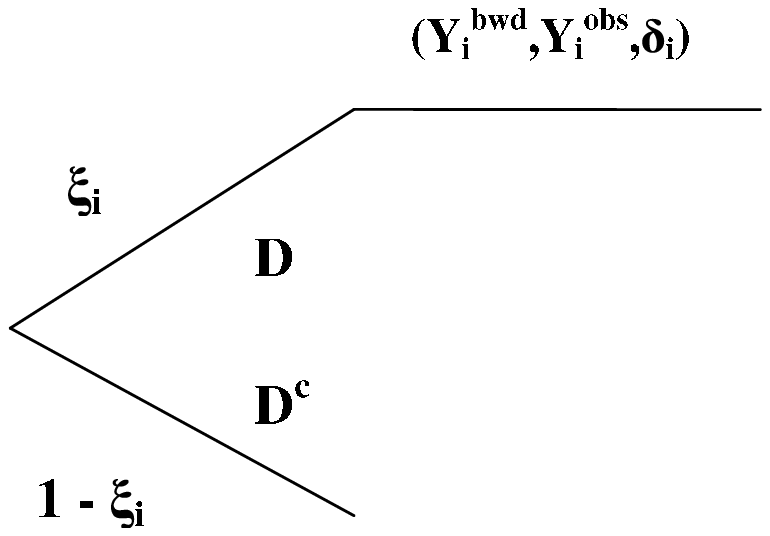}
\caption{Sampling scheme}
\label{Diagram1} \vspace{5mm}
\end{figure}

\noindent The full likelihood is:
\begin{equation} \label{likelihood}
L = \prod_{i=1}^{s}(1-P)^{1-\xi_{i}}\Bigg[P \Bigg
(\frac{dF(y_{i}^{bwd} + y_{i}^{fwd})} {\mu}\Bigg)^{\delta_{i}}
\Bigg(\int_{\omega \geq y_{i}^{bwd} + c_{i}}
\frac{dF(\omega)}{\mu}\Bigg)^{1-\delta_{i}}\Bigg]^{\xi_{i}},
\end{equation}
where $P=P(\xi = 1)$ is the time-independent point prevalence in the
population. For the derivation of a similar likelihood see
Asgharian and Wolfson (2005). As is readily seen, the above likelihood can be
factorized as
\begin{equation} \label{factor}
L = \Bigg [ \prod_{i=1}^{s} (1-P)^{1-\xi_{i}} P^{\xi_{i}} \Bigg] \
\Bigg[\prod_{i=1}^{s}\Bigg (\frac{dF(y_{i}^{bwd} + y_{i}^{fwd})}
{\mu}\Bigg)^{\delta_{i}} \Bigg(\int_{\omega \geq y_{i}^{bwd} +
c_{i}}
\frac{dF(\omega)}{\mu}\Bigg)^{1-\delta_{i}}\Bigg]^{\xi_{i}}\\
\end{equation}
Joint maximization of (\ref{likelihood}) with respect to $S$ and $P$
gives the NPMLE $(\hat{S}, \hat{P})$ and hence the NPMLE
$(\hat{\mu}, \hat{P})$ where $\hat{P} = N/s$ is the usual point
prevalence estimator of $P$. It follows, by invariance, that
$\hat{\lambda} = \hat{P}/\hat{\mu}$ is the unconditional NPMLE of
$\lambda$. It is seen that $\hat{\lambda}$, the MLE, is also the
natural ad hoc estimator derived from the relation $\lambda =
P/\mu$, by replacing $P$ and $\mu$ by their respective natural
estimators.

Wang (1991) derived the NPMLE $\hat{G}$ of $G$, the truncating distribution, by
conditioning on the observed backward recurrence times. Although,
under stationarity, $G$ is uniform, and this observation allows one
to informally assess stationarity, it does not lead to an estimate
of $\lambda$, since $\lambda$ is not uniquely determined by $G$.

\section{Interval estimation of the incidence rate} \label{CIlambda}

To derive an asymptotic confidence interval for $\lambda$ we begin with the
asymptotic properties of $(\hat{\lambda}, \hat{P})$ which in turn requires a
careful examination of the likelihood (\ref{likelihood}).
Identity (\ref{rep1}) of Lemma 1, though simple, plays a key role in the
derivation of an asymptotic confidence interval for $\lambda$ as it facilitates
the transferral of the asymptotic properties of $\hat{\mu}$ and $\hat{P}$ to
those of $\hat{\lambda}$.
\begin{lemma} \label{lem1}
Let $\lambda, P,$ and $\mu$ be respectively, the time-independent
underlying incidence rate, the time-independent point prevalence
and the mean duration of the disease. Let $\hat{\mu}$ and
$\hat{P}$ be the unconditional MLEs of $\mu$ and $P$
respectively. Define $\hat{\lambda}=\frac{\hat{P}}{\hat{\mu}}$, the
MLE of $\hat{\lambda}$. Then
\begin{equation} \label{rep1}
\hat{\lambda} - \lambda = \frac{1}{\hat{\mu} \mu} [\mu (\hat{P} -
P) - P(\hat{\mu} - \mu)] \quad .
\end{equation}
\end{lemma}
\begin{proof} The result follows immediately from the definitions of
$\lambda$ and $\hat{\lambda}$.
\end{proof}
Theorem \ref{AsgThm1} below, which draws on Lemma \ref{lem1},
essentially shows that $\hat{\lambda}$ is consistent and
asymptotically Normal. We state this result and provide the main steps of the
proof in the Appendix. Suppose
$\gamma = \sup\{t: F_{LB}(t) = 0 \}$, and $\tau = \inf\{t :
F_{LB}(t) = 1\}$. Then under mild conditions we have,
\begin{theorem} \label{AsgThm1}
Suppose $\gamma > 0$, $\tau < \infty $, and $\mu = \int_{0}^{\infty}
x dF(x) < \infty $. Then as $s \to \infty$ we have

\text{(a)} \hspace{.3 in} $ \mid \hat{\lambda} - \lambda \mid
~\stackrel{a.s.}{=}~  O \Big(\sqrt{\frac{\log \log s}{s}}\Big)$

\text{(b)} \hspace{.3 in} $ T = \sqrt{s} (\hat{\lambda} - \lambda)
\stackrel{\E{D}}{\to} \frac{Z_{1} - \lambda Z_{2}}{\mu}$,\\

\noindent where $Z_{1} \sim N(0, \lambda \mu[1 - \lambda \mu])$ and $Z_{2}
\sim N(0, \sigma^{2}_{\mu}/\lambda \mu)$ are independent,
\[
\sigma^{2}_{\mu} = \mu^{2} \int_{0}^{\infty} \int_{0}^{\infty}
\psi(u,v) d\Big(\frac{1}{u}\Big) d\Big(\frac{1}{v}\Big)~,
\]
and $\psi (u,v)$ is the covariance function of the limiting
process of $\hat{S}$.
\end{theorem}

\noindent The covariance function $\psi(u,v)$ has an intractable form (see Asgharian
et al. 2002 and Asgharian and Wolfson 2005). The asymptotic variance of $T$ has
consequently a rather complex form which precludes the possibility of its
direct estimation. Instead, we obtain a
confidence interval for $\lambda$ by bootstrapping $\hat{\lambda}$.

\section{Estimating the age-specific incidence rate} \label{age-specific}

For many diseases the incidence rate is age-dependent, and estimators of
age-specific incidence rates are almost always sought by epidemiologists.
Following the notation from Section \ref{notation}, let $\tau^{*}$
represent the calendar time of recruitment, let $X$ be the time
from onset to death, and $\tau_{o}$ be the calendar time of onset.
Let $D_{t}$ be the event of being diseased and alive at time $t$,
let $A_{o}$ be the age at onset, and $A_{t}$ the age at calendar
time $t$. We assume that the distribution of $X$ does not change
with calendar time, and that both $A_{o}$ and $A_{t}$ are discrete
random variables; the latter assumption can be relaxed to include
arbitrary random variables. Then,
\begin{align*}
 P(D_{\tau^{*}} | A_{o} = z)
&=
\int_{0}^{\tau^{*}} P(X \geq \tau^{*} - t , \tau_{o} = t \mid A_{o} = z) dt \\
&= \int_{0}^{\tau^{*}} P(X \geq \tau^{*} - t \mid \tau_{o} = t ,
A_{o} = z) d P_{\tau_{o} \mid A_{o}}(t \mid z) ~\cdot
\end{align*}
On the other hand, we have
\begin{align*}
d P_{\tau_{o} \mid A_{o}}(t \mid z)
&=
\frac{P(\tau_{o} \in (t, t+dt), A_{o}=z)}{P(A_{o}=z)} \\
&=
\frac{P(\tau_{o} \in (t, t+dt), A_{t} = z)}{P(A_{o}=z)} \\
&=
dP_{\tau_{o} \mid A_{t}}(t \mid z) \times \frac{P(A_{t} =
z)}{P(A_{o} = z)} ~\cdot
\end{align*}
We also note that
\[
P(X \geq \tau^{*} - t \mid \tau_{o} = t, A_{o} = z) = P(X \geq
\tau^{*} - t \mid A_{o} = z) = S_{z}(\tau^{*}-t) ~\cdot
\]
Having assumed that $dP_{\tau_{o} \mid A_{t}}(t \mid z)/dt = \lambda_{z}$ only
depends on $z$, we obtain
\[
P(D_{\tau^{*}} \mid A_{o} = z) = \Bigg[ \int_{0}^{\tau^{*}}
S_{z}(\tau^{*} - t) P(A_{t} = z) dt \Bigg] \frac{\lambda_z}{P(A_{o}
= z)} ~\cdot
\]
We thus find the age-specific incidence
\begin{equation} \label{a-s-t incidence}
\lambda_z = \frac{P(D_{\tau^{*}}, A_{o} = z)}{\int_{0}^{\tau^{*}}
S_{z}(\tau^{*} - t) P(A_{t} = z) dt }.
\end{equation}
It follows by invariance that the MLE of $\lambda_{z}$ is
\begin{equation} \label{a-s-t incidence estimate}
\hat{\lambda}_z = \frac{\hat{P}(D_{\tau^{*}}, A_{o} = z)}{\int_{0}^{\tau^{*}}
\hat{S}_{z}(\tau^{*} - t) P(A_{t} = z) dt }~,
\end{equation}
where $\hat{P}(D_{\tau^{*}}, A_{o}=z)$ is the observed proportion in
the recruited cohort who are diseased and with age-at-onset $z$.
Note that to find $\hat{S}_{z}$ we begin by restricting our
attention to the length-biased survival/censoring times of the
prevalent cases, whose onset occurred at age $z$. Then $\hat{S}_{z}$
is the MLE of $S_{z}$, based on these length-biased data, as derived
by Asgharian et al. (2002) and Asgharian and Wolfson (2005).

It is assumed that the population age distribution
$\{P(A_{t}=z)\}_{z}$, may be routinely obtained from census data.
Since census data are usually only updated every five years, a
reasonable assumption is that $P(A_{t}=z)$ is piecewise constant as
a function of $t$. However, as we shall see in Section
\ref{CSHAanalysis} it might be possible to make the even stronger
assumption that $P(A_{t}=z)=P(A=z)$, is roughly independent of $t$,
without affecting $\hat{\lambda}_{z}$ substantially. An alternative
which requires more intensive modeling, is to replace the step
function $P(A_{t}=z)$ by a smooth function of $t$. We suggest that
the extra effort would probably result in very small improvement if
any.

Since, in Section \ref{CSHAanalysis}, the population age
distribution is assumed to be constant we restrict our attention to
this case. Then equation (\ref{a-s-t incidence}) reduces to
\begin{equation} \label{a-s incidence}
\lambda_z = \frac{P(D_{\tau^{*}}, A_{o} = z)}{\mu_{z} ~ P(A=z)},
\end{equation}
where $P(A=z)$ is the proportion of subjects in age category $z$,
and $\mu_{z}$ represents the mean survival time in age category $z$.
The information contained in the observations, for the case of three
age categories ($z = 1, 2, 3$), may be illustrated through the
following tree diagram,
\begin{figure}[!h]
\centering
\includegraphics[scale=0.70]{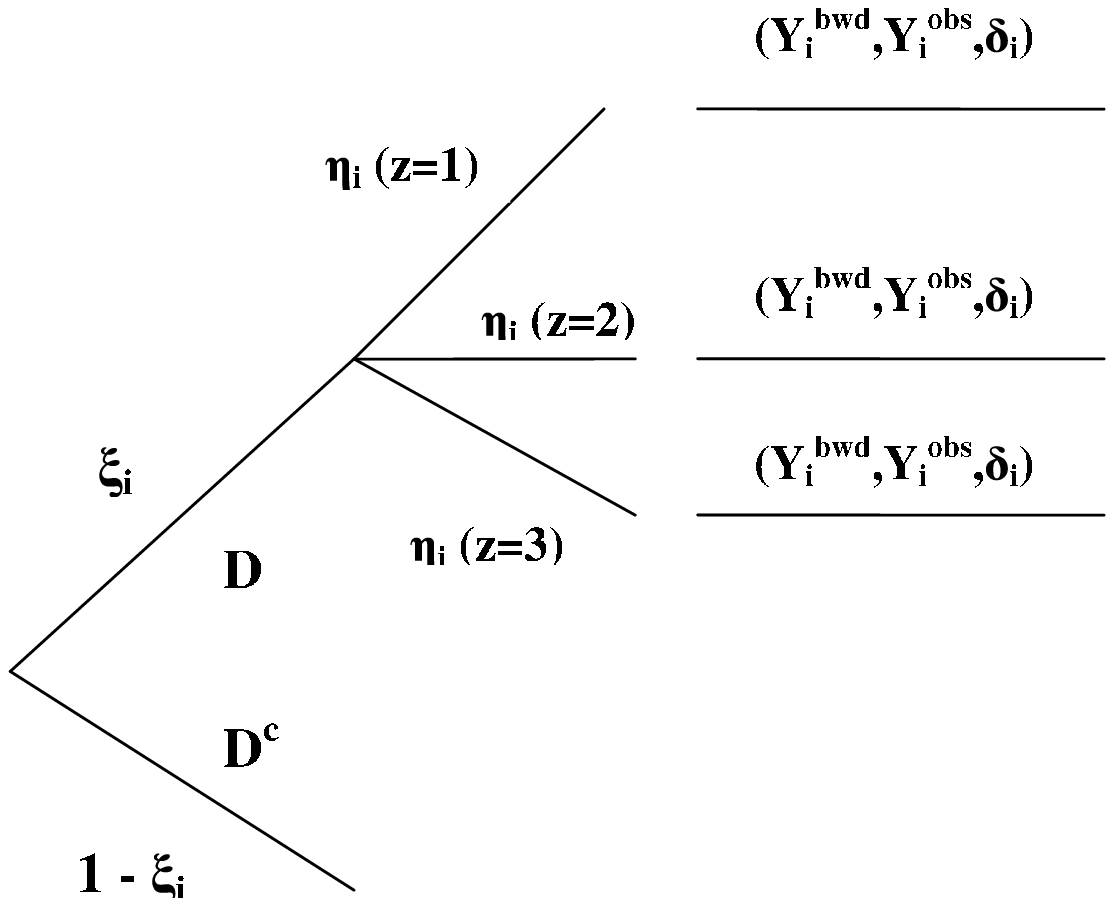}
\caption{Illustration of information contained in the observations}
\label{Diagram2} \vspace{3mm}
\end{figure}

\noindent where
\[
\eta_{z}(i) =
\begin{cases}
1, &\text{if $A_{o}=z$ for the i-th subject}, \\
0, &\text{Otherwise}.
\end{cases}
\vspace{3mm}
\]

\noindent The full likelihood, for the general case $z = 1, 2,...,l$
is,
\begin{equation} \label{age-specific likelihood}
\hspace{-7mm} L_{a} = \Bigg [ \prod_{i=1}^{s} (1-P)^{1-\xi_{i}}
P^{\xi_{i}} \Bigg] \ \Bigg[\prod_{i=1}^{s} \Big \{ \Big
(\frac{dF(y_{i}^{bwd} + y_{i}^{fwd})} {\mu}\Big)^{\delta_{i}}
\Big(\int_{\omega \geq y_{i}^{bwd} + c_{i}}
\frac{dF(\omega)}{\mu}\Big)^{1-\delta_{i}} \prod_{z=1}^{l}
P(A_{o}=z)^{\eta_{z}(i)} \Big \}^{\xi_{i}} \Bigg]. \vspace{5mm}
\end{equation}

\noindent Using equation (\ref{a-s incidence}),
\begin{equation} \label{asincidenceestimate}
\hat{\lambda}_{z} = \frac{\hat{P}(D_{\tau^{*}},
A_{o}=z)}{\hat{\mu}_{z} P(A=z)}~,
\end{equation}
is the MLE of $\lambda_{z}$, where $\hat{\mu}_{z}$ is the MLE of
$\mu_{z}$ derived from $\hat{S}_{z}$.

\section{Estimating the incidence rate of dementia} \label{CSHAanalysis}

In 1991, 10,263 elderly Canadians (65 years or older), living at
home or in an institution, were screened for dementia (CSHA
working group 1994). This phase of the study was known as CSHA-1.
At the time of CSHA-1, 821 subjects were classified as having
either possible Alzheimer's disease, probable Alzheimer's disease,
or vascular dementia. Henceforth, by the term dementia we mean
having exactly one of these three conditions since they constitute
the vast majority of dementias. The approximate dates of onset
were derived in a hierarchical fashion from the answers to three
questions (Wolfson et al. 2001). In 1996, the second phase of the
study, CSHA-2, was completed. CSHA-2 included the ascertainment of
the date of death or right censoring for those cases identified at
CSHA-1. These are the data upon which we shall base our estimates
of the overall and age-specific incident rates of dementia.
However, additional data were, in fact, collected as part of the
CSHA with the goal of estimating the age-specific incidence rates
of dementia among elderly Canadians. The subjects who were deemed
not to have dementia at CSHA-1 were re-evaluated for dementia at
CSHA-2. Assuming that these incidence rates had remained constant,
they were estimated using the incident cases observed between
CSHA-1 and CSHA-2. There were nevertheless, difficulties with
these ``incident'' data since it could not be ascertained with
certainty whether those who had died between CSHA-1 and CSHA-2 had
become incident cases with dementia. In this paper we, therefore,
re-estimated the incidence rates without relying on the
``incident'' cases that occurred between CSHA-1 and CSHA-2. The
assumption of a roughly constant incidence rate for dementia has been previously
checked in several ways and has been deemed to be reasonable (see Asgharian
et al. 2002, Asgharian et al. 2006 and Addona and Wolfson 2006).

\subsection{Estimating the overall incidence rate of dementia}
The NPMLE, $\hat{S}(x)$, yields $\hat{\mu} \approx$
4.75 years or 57 months. Since the CSHA data did not constitute a
random sample of all subjects over the age of 65 in Canada, we used
the age-standardized prevalence estimate instead of simply, $\hat{P}
= \frac{N}{s}$~. This gives an estimate for $P$ of 0.066 (CSHA
working group 1994), which leads to a point estimate, $\hat{\lambda}
=$ 0.0139, or 13.9 per 1,000 person-years. To obtain an interval
estimate for $\lambda$, we followed the bootstrap procedure
and sampled with replacement from the 10,263 screened subjects to obtain
10,000 bootstrap samples of the same size. We obtained a confidence interval
for $\lambda$ of $[12.52,15.28]$ cases per 1,000 person-years.

\subsection{Estimating the age-specific incidence rate of dementia}

\begin{figure}[!h]
\centering
\includegraphics[scale=0.70]{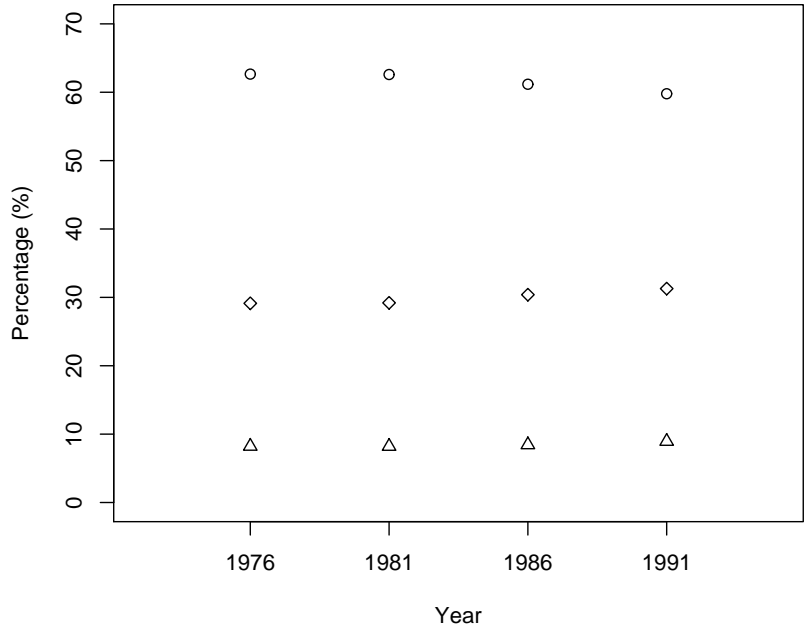}
\caption{Percentage of Canadian population aged 65 and older in
65-74 (circles), 75-84 (diamonds), and 85+ (triangles) age groups}
\label{ageprop} \vspace{5mm}
\end{figure}

Three age groups were considered for the CSHA data: 65-74, 75-84,
and 85+ years old. The 821 cases of dementia were subdivided as
follows: 164 had onset between 65 and 74 years old, 381 had onset
between 75 and 84 years old, and 276 were 85 or older when they had
onset. The estimated mean survival times in years were 7.97, 5.16,
and 3.50, for the 65-74, 75-84, and 85+ groups, respectively. To use
equation (\ref{asincidenceestimate}), we require an approximately
stable age distribution over the period covering the onset times. We
consulted data from four Canadian censuses covering 1976-1991 to
assess this assumption. Figure \ref{ageprop} shows the progression
of the percentage of the Canadian population aged 65 and older in
each of the three age groups (Statistics Canada 2006). Using data
from the 1976, 1981, 1986, and 1991 censuses, we also computed some
measures of variability for the percentage in each of the three age
groups. These are presented in Table 1.

\begin{table}
\center
\begin{tabular}{|c|c|c|c|} \hline
  ~~~~\textbf{Age Group}~~~~& \textbf{Range} & \textbf{SD} & \textbf{CV} \\ \hline
  \textbf{65-74} & ~~59.8 - 62.7\%~~ & ~~1.36\%~~& ~~0.022~~\\ \hline
  \textbf{75-84} & ~~29.1 - 31.3\%~~ & ~~1.03\%~~& ~~0.034~~\\ \hline
  \textbf{85+} & ~~~8.2 - 8.9\%~~~   & ~~0.34\%~~& ~~0.040~~\\ \hline
\end{tabular}
\vspace{.1 in} \caption{Range, standard deviation, and coefficient
of variation for the percentage of Canadians (aged 65 and older)
in each age group}
\end{table}

\noindent Having verified that the age distribution is roughly
stable for this time period, we proceeded with the age-specific
incidence estimation using the 1991 census data. Amongst those 65
years or older in 1991, 59.8\% were in the 65-74 group, 31.3\%
were in the 75-84 group, and 8.9\% were in the 85+ group
(Statistics Canada 2006). The resulting age-specific incidence
rate estimates are presented in Table 2. In 1976, amongst those 65
years or older, 62.7\% were in the 65-74 group, 29.1\% were in the
75-84 group, and 8.2\% were in the 85+ group (Statistics Canada
2006). We also estimated the age-specific incidence rates based on
the census age distribution data from 1976 to take into account
small changes in the population age distribution that might have
occurred over the period from 1976 to 1991. When the age
distribution changes, this approach provides a simple framework
for investigating robustness of the age-specific incidence rate
estimator to departures from the assumption of constancy of the
age distribution. For comparative purposes, the age-specific
incidence rate estimates based on the 1976 census data are also
given in Table 2.
\begin{table}
\center
\begin{tabular}{|c|c|c|c|c|} \hline
\textbf{Age Group}&~\textbf{$\hat{\lambda}_{z}$ (1991)}~&
\textbf{95\% CI (1991)}&~\textbf{$\hat{\lambda}_{z}$
(1976)}~&\textbf{95\% CI (1976)}\\\hline
  \textbf{65-74} & 3.35 & [2.72 , 3.99] & 3.20 & [2.58 , 3.82]\\ \hline
  \textbf{75-84}& 22.99 & [19.92 , 26.04] & 24.69 & [21.26 , 28.13]\\ \hline
  \textbf{85+}  & 85.86 & [70.52 , 101.20] & 93.39 & [77.00 , 109.77]\\ \hline
\end{tabular}
\vspace{.1 in} \caption{Age-specific incidence rate estimates per
1,000 person-years using 1991 and 1976 Canadian census data}
\end{table}

\subsection{Discussion of dementia incidence rate estimates} \label{sect63}

Subjects were not prospectively monitored between CSHA-1 and
CSHA-2. It was thus difficult to ascertain whether those who had
died in this time period had had onset of Alzheimer's disease
(possible or probable). As a result, incidence rates of
Alzheimer's disease reported from the CSHA were underestimates of
the true incidence rates amongst elderly Canadians since they were
based only on subjects who survived until the end of CSHA-2 in
1996. The CSHA incidence rate estimates of Alzheimer's disease were 7.4
and 5.9 per 1,000 person-years for women and men respectively
(CSHA working group 2000), giving a crude estimated incidence rate for men
and women combined of 6.7 per 1000 person years. Using the CSHA data,
H\'{e}bert et al.(2000) estimated the incidence rate of vascular dementia to be
3.79 per 1,000 person-years. Therefore the CSHA overall (under-) estimated rate
for dementia was approximately 6.7 + 3.79 = 10.49 per 1000 person years. Direct
comparison with our results is difficult. Our overall estimate (for
possible or probable Alzheimer's, or vascular dementia) of 13.9
per 1,000 person-years seems to be consistent with these previous
estimates obtained from the CSHA. An analogous comparison of the
age-specific incidence estimates reveals that they too are
consistent with those already obtained from the CSHA. Note that
the slightly different point estimates for the age-specific
incidence rates, particularly in the 85+ category, should not be
interpreted as meaning that incidence rates declined or increased
between 1976 and 1991. They simply provide a range of possible
values for the estimate depending on what is taken as the
population age distribution.

\section{Concluding remarks} \label{remarks}

Simulations, whose results are not reported in this paper, suggest that our methods
work well for moderate sample sizes; the asymptotic distribution of the estimated
incidence rates were close to Normal, the point estimates were close to their true
values and the confidence intervals reasonably narrow for a range of parameter choices.

Our estimator of the incidence rate depends on the estimator $\hat{S}$, for the survival
function $S$. We propose that the most efficient estimator for $S$, under the assumption
of a constant incidence rate, should be used. The estimator, $\hat{S}$, used in this paper
is more efficient than the well-known estimator of $S$ for general left truncation data
(Wang 1991) which does not invoke stationarity of the incidence process (Asgharian et al. 2002).
Indeed, it is possible to show that the estimators we present for the incidence rates
(overall and age-specific) are asymptotically most efficient. This follows from Asgharian
and Wolfson (2005, Theorem 3) and Van der Vaart (1998, Theorem 25.47, page 387).

If the largest observed failure time is censored, $\hat{S}$ is left undefined beyond this
point by most authors. Consequently $\hat{\mu}$ is not well-defined in this situation.
Fortunately, in our example, the largest survival time is a true failure time. Ad hoc
``fixes'' are available, but produce biased estimators.

The CSHA data used for illustration is based on an initial cohort of
10,263 subjects obtained as a stratified cluster sample whereby a
fixed number of institutionalized (about 10\%) and
non-institutionalized (about 90\%) subjects were sampled. In
addition, those over 85 years old were over-sampled. We do not take
into account the sampling scheme in our estimated incidence rates or
in the asymptotic distributions of our estimators. To do so requires
development of new theory allowing for within cluster dependence,
which is a topic for further study and is not directly pertinent to
our methods.

\section*{Acknowledgments}

This research was supported in part by the Natural Sciences and
Engineering Research Council of Canada (NSERC) and Le Fonds
Qu\'{e}b\'{e}cois de la recherche sur la nature et les technologies
(FQRNT). The data reported in this article were collected as part of
the CSHA. The core study was funded by the Seniors' Independence
Research Program, through the National Health Research and
Development Program (NHRDP) of Health Canada Project
6606-3954-MC(S). Additional funding was provided by Pfizer Canada
Incorporated through the Medical Research Council/Pharmaceutical
Manufacturers Association of Canada Health Activity Program, NHRDP
Project 6603-1417-302(R), Bayer Incorporated, and the British
Columbia Health Research Foundation Projects 38 (93-2) and 34
(96-1). The study was coordinated through the University of Ottawa
and the Division of Aging and Seniors, Health Canada.

We would also like to thank the referees and Associate Editor for
their useful comments and suggestions which helped greatly enhance
our paper.

\section*{Appendix}

We provide a road map of the proof of Theorem \ref{AsgThm1} and give further
details about steps (iv) and (v) below.  Road map of the proof:
\newline
(i) Establish the asymptotic behavior of $\hat{F}_{LB}$.
\newline
(ii) Establish the asymptotic behavior of $\hat{F}$ using (i).
\newline
(iii) Establish the asymptotic behavior of $\hat{\mu}$ using (ii).
\newline
(iv) Establish the independence of $\hat{\mu}$ and $\hat{P}$.
\newline
(v) Establish the asymptotic behavior of $\hat{\lambda}$ using
(\ref{rep1}), (iii), and (iv).
\newline
The derivation of (i) is similar to its counterpart given by
Asgharian and Wolfson (2005), and those of (ii) and (iii) are
similar to their counterparts in Asgharian et al. (2002). The
expressions, however, are slightly different in view of the
different sampling scheme under consideration here. We therefore
sketch the proof of steps (iv) and (v).

\hspace{-.4 in} {\bf Step(iv):} \vspace{.1 in}
\newline
In this step we justify the independence of $\hat{P}$ and
$\hat{F}$, and hence of $\hat{P}$ and $\hat{\mu}$. This
independence is suggested by the likelihood factorization
(\ref{factor}). Theorem \ref{Thm3} shows that this is in fact the
case. First, observe that it follows from (\ref{likelihood}) that
$\hat{P} = \frac{N}{s}$ is the NPMLE of $P$. We have
\begin{equation} \label{Pi}
\Pi_{s} = \sqrt{s}(\hat{P} - P) \stackrel{\E{D}}{\to} N \Big
(0,P(1-P) \Big) \quad .
\end{equation}

\begin{theorem} \label{Thm3}
Let $\Bbb{R} \times D_{0}[0, t]$ be endowed with the topology
induced by
\[
\parallel (a, x) \parallel = \mid a \mid +
\sup_{s \in [0, t]} \mid x(s) \mid \ \quad .
\]
Under the assumptions of Theorem 1
\[
(\Pi_{s}, U_{N}) \stackrel{\E{D}}{\to} (W, U) \qquad in \quad
\Bbb{R} \times D_{0}[0, t] \ ,
\]
where $U$ is given in Theorem 1 and $W \sim N(0, P(1-P))$ is
independent of $U$.
\end{theorem}

\begin{proof} It follows from Theorem 7.2.1 and Lemma 7.2.1 of Cs$\ddot{o}$rgo
and Revesz (1981) that
\[
\parallel (\Pi_{s}, U_{N}) - (\Pi_{s}, U_{[sP]}) \parallel \stackrel{P}
{\longrightarrow} 0 \quad \text{as} \quad s \to \infty \
\]
It remains to show that $\Pi_{s}$ and $U_{[sP]}$ are independent.
This follows from the fact that $\hat{P}$ is a partial ancillary
for $F_{LB}$, while $\{(Y_{i}^{bwd}, Y_{i}^{obs}, \delta_{i}),
\quad i=1,2, \cdots, [sP]\}$ is partially sufficient for $F_{LB}$.
\end{proof}
\hspace{-.4 in} {\bf Step(v):} \vspace{.1 in}
\newline
In this final step we combine the results of steps (i) through
(iii), in Theorem \ref{AsgThm1}, to yield the asymptotic behaviour
of $\hat{\lambda}$.

{\it Proof of Theorem \ref{AsgThm1}.} Part \emph{(a)} follows from
Lemma \ref{lem1}, part (b) of Theorem 1 of Asgharian et al. (2002), and the asymptotic
properties of the sample proportion in Binomial sampling. Part (b)
follows from part (c) of Theorem 1 of Asgharian et al. (2002), (\ref{Pi}), and the
identity (\ref{rep1}). \hspace{4.1 in} $\Box$

\vspace{1cm}

\section*{References}

\hspace{-.4 in}
1. Addona, V. and Wolfson, D. B. (2006). A formal test for
the stationarity of the incidence rate using data from a
prevalent cohort study with follow-up. {\it Lifetime Data
Analysis} {\bf 12}, No. 3, 267-284.
\newline
2. Alioum A., Commenges D., Thi\'{e}baut R., and Dabis F.
(2005). A multistate approach for estimating the incidence
of human immunodeficiency virus by using data from a
prevalent cohort study. {\it Applied Statistics} {\bf 54}, Part 4,
739-752.
\newline
3. Asgharian, M., M'Lan, C.E. and Wolfson, D. B. (2002).
Length-biased sampling with right censoring: an
unconditional approach. {\it Journal of the American Statistical
Association} {\bf 97}, No.457, 201-209.
\newline
4. Asgharian, M., and Wolfson, D.B. (2005). Asymptotic
behaviour of the unconditional NPMLE of the length-biased
survivor function from right censored prevalent cohort data.
{\it The Annals of Statistics} {\bf 33}, No.5, 2109-2131.
\newline
5. Asgharian, M., Wolfson, D. W. and Zhang, X. (2006).
Checking stationarity of the incidence rate using
prevalent cohort survival data. {\it Statistics in Medicine} {\bf
25}, 1751-1767.
\newline
6. Biering-Sorensen, F., and Hilden, J. (1984). Reproducibility of
the history of low-back trouble. {\it Spine} {\bf 9}, No.3, 280-286.
\newline
7. Cox, D.R. (1969). Some sampling problems in technology. In {\it
New Developments in Survey Sampling}. Edited by Johnson and Smith.
Wiley, 506-527.
\newline
8. Csorgo, M. and Revesz (1981). {\it Strong Approximation
in Probability and Statistics}. Academic Press, New York.
\newline
9. CSHA working group, (1994). Canadian study of health and aging:
study methods and prevalence of dementia. {\it Journal of the
Canadian Medical Association} {\bf 150}, 899-913.
\newline
10. CSHA working group (2000). The incidence of dementia in Canada.
{\it Neurology} {\bf 55}, 66-73.
\newline
11. Diamond, I.D., and McDonald, J.W. (1991). Analysis of
current-status data. In {\it Demographic Applications
of Event History Analysis} Edited by Trussell, J.,
Hankinson, R., and Tilton, J. chapter 12. Oxford: Oxford
University Press.
\newline
12. Folnegovic, Z. and Folnegovic-Smalc, V. (1992).
Schizophrenia in Croatia: interregional differences in
prevalence and a comment on constant incidence. {\it
Journal of Epidemiology and Community Health} {\bf 46}, 248-255.
\newline
13. H\'{e}bert, R., Lindsay, J., Verreault, R., Rockwood,
K., Hill, G., and Dubois, M-F.(2000). Vascular dementia:
Incidence and risk factors in the Canadian Study of Health
and Aging. {\it Stroke} {\bf 31}, 1487-1493.
\newline
14. Jemal, A., Murray, T., Ward, E., Samuels, A., Tiwari, R.C.,
Ghafoor, A., Feuer, E.J., and Thun, M.J. (2005). Cancer Statistics,
2005. {\it CA: A Cancer Journal for Clinicians} {\bf 55}, 10-30.
\newline
15. Keiding, N. (2006). Event history analysis and the
cross-section. {\it Statistics in Medicine} 25, 2343-2364.
\newline
16. Keiding, N. (1991). Age-specific incidence and prevalence: A
statistical perspective (with discussion). {\it Journal of the Royal
Statistical Society, Series A} {\bf 154}, No.3, 371-412.
\newline
17. Keiding, N., Holst, C. and Green, A. (1989). Retrospective
estimation of diabetes incidence from information in a current
prevalent population and historical mortality. {\it American Journal
of Epidemiology} {\bf 130}, 588-600.
\newline
18. Neyman, J. (1955). Statistics-servant of all sciences. {\it
Science}, {\bf Vol.122}, no. 3166, 401-406.
\newline
19. Ogata, Y., Katsura, K., Keiding, N., Holst, C. and Green, A.
(2000). Empirical Bayes age-period-cohort analysis of retrospective
incidence data. {\it Scandinavian Journal of Statistics} {\bf 27},
415-432.
\newline
20. Patil, G.P. and Rao, C.R. (1978). Weighted distributions and
sized-biased sampling with applications to wildlife populations and
human families. {\it Biometrics} {\bf 34}, 179-189.
\newline
21. Sorenson, E.J., Stalker, A.P., Kurland, L.T., and Windebank,
A.J. (2002). Amyotrophic lateral sclerosis in Olmsted County,
Minnesota, 1925 to 1998. {\it Neurology} {\bf 59}, 280-282.
\newline
22. Statistics Canada (2006) Census of Population, Statistics Canada
catalogue no. 97- 551-XCB2006005: Age Groups and Sex for the
Population of Canada, Provinces and Territories, 1921 to 2006
Censuses. Ottawa: Statistics Canada.
\newline
23. Van der Vaart, A. W.(1998). \emph{Asymptotic Statistics}.
Cambridge Series in Statistics and Probabilistic Mathematics.
Cambridge University Press, Cambridge, UK.
\newline
24. Vardi, Y. (1982). Nonparametric estimation in the presence of
length bias. {\it The Annals of Statistics} {\bf 10}, 616-620.
\newline
25. Vardi, Y. (1985). Empirical distributions in selection bias
models. {\it The Annals of Statistics} {\bf 13}, 178-205.
\newline
26. Vardi, Y. (1989). Multiplicative censoring, renewal processes,
deconvolution and decreasing density: nonparametric estimation. {\it
Biometrika} {\bf 76}, No.4, 751-761.
\newline
27. Wang, M-C. (1991). Nonparametric estimation from cross-sectional
survival data. {\it Journal of the American Statistical Association}
{\bf 86}, No.413, 130-143.
\newline
28. Wicksell, S.D. (1925). The Corpuscle Problem: A Mathematical
Study of a Biometric Problem. {\it Biometrika} {\bf 17}, No.1,
84-99.
\newline
29. Wolfson, C., Wolfson, D.B., Asgharian, M., M'Lan, C.E.,
{\O}stbye, T., Rockwood, K. and Hogan, D.B.(2001). A reevaluation of
the duration of survival after the onset of dementia. {\it New
England Journal of Medicine} {\bf 344}, No.15, 1111-1116.
\end{document}